\documentclass{article} 

\usepackage[utf8]{inputenc} 
\usepackage[english]{babel} 
\usepackage{amsmath}
\usepackage{amssymb}
\usepackage{txfonts}
\usepackage{mathdots}
\usepackage[classicReIm]{kpfonts}
\usepackage{graphicx}

\begin{document}

\title{Kalman Filter in the Problem of the Exchange and the Inflation Rates Adequacy To Determining Factors\footnote{\ The\ work\ was\ supported\ in\ part\ by\ The\ National\ Academy\ of\ Sciences\ of\ Ukraine\ \\ (project\ No.\ 0118U003196)\ and\ by\ the\ Program\ of\ Fundamental\ Research\ of\ the\ Department\ \\of\ Physics\ and\ Astronomy\ of\ the\ National\ Academy\ of\ Sciences\ of\ Ukraine\ \\(project\ No.\ 0117U000240).}
}
\author{N. S. Gonchar, W. H. Kozyrski, A. S. Zhokhin, O. P. Dovzhyk\\
Bogolyubov Institute for Theoretical Physics of NAS of Ukraine.}
\date{}
	
\maketitle
\noindent \textbf{}

\noindent 


\begin{abstract}
Using introduced concept of the exchange and inflation rates adequacy, the relevance of them to the determining factors is found. We established close positive relation between hryvnia / dollar exchange and inflation rates, fiscal deficit, price level of energy sources, and money supply. On this basis, we give proposals for state macroeconomic policy to stabilize Ukrainian economy.
\end{abstract}

\noindent \textbf{1.~Introduction}

\textbf{}

The first part of the paper contains a study of the influence of determining factors on the hryvnia / dollar exchange rate and the level of inflation in the Ukrainian economy. The most important factors that significantly affect the hryvnia / dollar exchange rate and inflation rate are payment balance, trade balance, energy prices, state budget deficit, money supply, and refinancing rate of the National Bank. In noncompetitive economies, the formation of the exchange rate and the level of inflation occurs under the influence of monopolies and taxation systems. The question arises whether the hryvnia / dollar course and the inflation rate are relevant to the factors they depend on.

Here we introduce the concept of an adequate hryvnia / dollar exchange rate and the level of inflation to the factors they depend on. On this basis, a stochastic dynamic system of an adequate hryvnia / dollar and inflation rate was formulated and investigated. In the constructed stochastic model of an adequate hryvnia / dollar and inflation rate, these values {}{}are not observable. Observed factors are those they depend on. The assumptions made allow us to find the best mean squared estimate using Kalman's method [1,2]. It has been established that during 2012-2013 the exchange rate was artificially kept from devaluation that resulted in a sharp depreciation of the national currency in 2014.  

The purpose of the theoretical exchange rate model is to understand the economic mechanism of exchange rate management and the relationship of the latter to other economic variables. The construction of an adequate model of the exchange rate of currencies requires examining the empirical patterns that characterize the behavior of exchange rates and other related variables at a floating exchange rate [3].

The results of studying the dynamics of the exchange rates of industrial countries in the 1970s, cited in [3], can be summarized as follows. Statistics of exchange rate logarithms showed that they change in time approximately as a random walk with or without a drift implying the inability to predict the structure of monthly changes in exchange rates. Only a small part of such changes can be predicted by the market measurement of forward discounts and premiums. Analysis of simultaneous correlations between the movement of the spot and forward rates (up to one year) shows their unidirectional change by approximately the same amount, especially for rather large changes. The correlation of the exchange rate change with the change in the ratio of price indices at small time intervals was close to zero.

Positive serial correlations of monthly changes in the ratio of price indices did not correspond to serial correlations of monthly changes in exchange rates. The cumulative divergences of relative purchasing power parities and exchange rates among the major industrial countries ranged within 10\%.

 Monthly changes in the nominal exchange rate are closely correlated with monthly changes in real exchange rates, and cumulative changes in real exchange rates during the year were quite large. By the real exchange rate one understands the price of a unit of foreign currency divided by the ratio of the national price index to the foreign one.

 There is no strong systematic relation between changes in the nominal and real rates and current account balances. Changes in nominal and real rates are not too tied to the rate of money supply, except, perhaps, for economies with high inflation [3].

  With a sufficiently low demand for the determining group of goods, the economic system may fall into recession [4, 5, 6, 7]. At the macroeconomic level, this is accompanied by a devaluation of the national currency, inflation, a fall in asset prices. The phenomenon was called the breakdown of exchange mechanism. It is possible to reveal it in advance from the statistical information on the structure of production, outputs, foreign economic relations.

There are many factors destabilizing economy, however, all they are only a trigger to scrap the exchange mechanism. The hidden cause of exchange mechanism breakdown is the quality of equilibrium state that can be determined by the structural characteristics of the economy [4, 5, 6, 7]. Ibid., the model of economic equilibrium is examined, the quality of the state of economic equilibrium is investigated, and the mechanism of recession is indicated.

How does monetary policy affect the recession in the economy? Here, on the example of the Ukrainian economy, the question is raised about overcoming inflation and the role of the hryvnia / dollar exchange rate in it has been clarified. First, we study the influence of monetary policy on the exchange rate in nonlinear regression models and establish the equation of monetary circulation. The description of the hryvnia / dollar exchange rate by a random process satisfying the difference stochastic equation confirms that the logarithm of the exchange rate is a random walk without a drift. On this basis, the equation of money circulation in 2012 - 2014 was found. We showed an increase in the rate of money circulation in that period. This indicates a decline in the standard of living. We discovered a close positive correlation between the hryvnia / dollar exchange rate and the level of inflation, a deficit in the state budget, a price level for energy carriers, and money supply. The recommendations on the macroeconomic policy of the state to stabilize the economy have been worked out. Finding out what the state's macroeconomic policy should be to reduce inflation, we established a close positive correlation between the inflation level and the hryvnia / dollar exchange rate, a deficit in the state budget, and a trade balance deficit. The latter shows that it is impossible to stabilize inflation without stabilizing the exchange rate.

\noindent 

\noindent \textbf{2.}~\textbf{The Adequacy Of the Exchange And the Inflation Rates To Determining Factors}

\noindent 

\noindent \textbf{     }Formation of the exchange and inflation rates in non-competitive economies may occur inadequately to the economically determining factors. This is due to the influence of monopoly on economic processes and the deformed taxation system, which contribute to inflationary processes and inadequate changes in the hryvnia / dollar rate. Using introduced concept of adequacy of the exchange inflation rates, we find out their relevance to the determining factors.

We believe that the hryvnia / dollar and the inflation rate are determined by the factors such as energy prices, current balance of payments, trade balance, money supply, total foreign currency reserves, interest rate refinancing, and GDP.

The complexity of the problem lies in the fact that there is no unambiguous stochastic dynamic evolution of the economy. The main assumption is that the course of hryvnia / dollar and the rate of inflation are determined mainly by a certain set of factors and is an unobservable process whose estimate should be found.

\textit{Formulation of the problem}. The evolution of factors $Y_{i} ,i=\overline{1,N}$is given that can affect the inflation and the exchange rates, as well as the evolution of the inflation rate and the exchange rate of hryvnia / dollar $x_{i} \; ,\; i=\overline{1,N}$, where the dimensionality of the vector of factors is $n$, and the dimensionality of the vector $x$ is equal to 2. One must clarify the adequacy of the influence of factors on the inflation and exchange rates.
\\ \\
We'll solve the problem by constructing unobservable random process $x_{k} $ being adequate inflation rate and exchange rate within $k$-th period that evolves by the law
\begin{equation} \label{GrindEQ__1_} 
\bar{x}{}_{k+1} =F_{k,k+1} \bar{x}_{k} +w_{k} ,\quad k=0,N-1,                                                                                             
\end{equation} 
\begin{equation} \label{GrindEQ__2_} 
Y_{k+1} =H_{k} \bar{x}_{k} +\nu _{k} ,\quad k=\overline{0,N-1}   ,                                                                                              
\end{equation} 
considering $w_{k} $ and $\nu _{k} $ to be Gauss white noises, i.e.,  $E\; w_{k}^{i} \nu _{s}^{j} =0$, where                                                    
\[E\; w_{k}^{i} \nu _{s}^{j} =\delta _{ks} \delta _{ij} a^{i} ,\quad a^{i} >0,\; \; i=1,{\kern 1pt} \, 2,  E\; \nu _{k}^{i} \nu _{s}^{j} =\delta _{ks} \delta _{ij} b^{i} ,\quad b^{i} >0,\, \; i=\overline{1,N},\] 
with zero mean  $E\; w_{k} =0,\; \; E\; \nu _{k} =0$.  $F_{k,k+1} ={\rm E} $is unit $2\times 2$ matrix.

We determine the matrix $H_{k} $ from the condition of minimum for the functional 
\begin{equation} \label{GrindEQ__3_} 
\min \; \sum _{i=1}^{n}\{ [y_{i}^{k}  -h_{i1}^{k} x_{1}^{k} -h_{i2}^{k} x_{2}^{k} ]^{2} +[y_{i}^{k+1} -h_{i1}^{k} x_{1}^{k+1} -h_{i2}^{k} x_{2}^{k+1} ]^{2} \}  ,                                                  
\end{equation} 
where  $H_{k} =\left|h_{i1}^{k} ,h_{i2}^{k} \right|_{i=1}^{n} \; ,\; Y_{i}^{k} =\{ y_{i}^{k} \} _{i=1}^{n} \; ,\; k=1,N$.

At the beginning time moment

\begin{equation} \label{GrindEQ__4_} 
\hat{x}_{0} =E\; \bar{x}_{0}  ,                                                                                    
\end{equation} 
\begin{equation} \label{GrindEQ__5_} 
P_{0} =E\, \; (\bar{x}_{0} -E\, \bar{x}_{0} )\, (\bar{x}_{0} -E\, \bar{x}_{0} )^{T} .                                                   
\end{equation}

We describe the evolution of the state estimate by the formula [2]

\begin{equation} \label{GrindEQ__6_} 
\hat{x}_{{}_{k} }^{-} =F_{k,k-1} \hat{x}_{{}_{k-1} }^{-}    .                                                                           
\end{equation}

Errors matrix within the time period  $[k-1\, ,\; \, k]$ is given by the expression

\begin{equation} \label{GrindEQ__7_} 
P_{k}^{-} =F_{k,k-1} P_{k-1} F_{k,k-1}^{T} +Q_{k-1}  .                                                           
\end{equation} 

Kalman matrix

\[G_{k} =P_{k}^{-} \, H_{k}^{T} \, [H_{k} \, P_{k}^{-} \, H_{k}^{T} +R_{k} ]^{-1} \]

and                                     $\hat{x}_{k} =\hat{x}_{{}_{k} }^{-} +G_{k} \, (Y_{k+1} -H_{k} \hat{x}_{k}^{-} )$.

Updated bug matrix is

\[P_{k} =(E-G_{k} \, H_{k} )\, P_{k}^{-}  .\]

We suppose that $F_{k,k+1} ={\rm E} $, where  ${\rm E} $ is unit $2\times 2$matrix.

In the \textbf{Table 1}, we show the quarterly change in the inflation rate and the hryvnia / dollar rate.

Here, the first and third columns contain inflation dynamics in respect to previous year data. The second and the fourth columns present the cost \$ 100 in UAH.
\vskip 5mm
\textbf{Table 1.} Forecast of the exchange and inflation rates

\vskip 2mm

\begin{tabular}{|p{0.6in}|p{0.7in}|p{0.6in}|p{0.7in}|} \hline 
\multicolumn{2}{|p{1in}|}{Statistical data} & \multicolumn{2}{|p{1.3in}|}{Forecast} \\ \hline 
2.82 & 798.90 &  &  \\ \hline 
1.89 & 799.00 &  &  \\ \hline 
1.70 & 799.30 & 1.20 & 802.74 \\ \hline 
1.89 & 799.30 & 1.79 & 821.56 \\ \hline 
1.68 & 799.30 & 2.18 & 789.61 \\ \hline 
1.39 & 799.30 & 1.88 & 854.32 \\ \hline 
1.55 & 799.30 & 1.28 & 817.80 \\ \hline 
1.23 & 799.30 & 1.57 & 837.96 \\ \hline 
1.41 & 885.60 & 1.59 & 826.17 \\ \hline 
2.05 & 1169.35 & 1.38 & 884.33 \\ \hline 
1.78 & 1257.59 & 2.55 & 1393.60 \\ \hline 
1.25 & 1442.13 & 1.63 & 1246.11 \\ \hline 
\end{tabular}

\vskip 2mm
\textbf{Source:} Statistical data were presented by the Ukrainian National Bank and forecast magnitudes were obtained within the procedure described.
\vskip 4mm
As a result of model study, we reveal the fact of artificial refraining of exchange rate in the years 2012-2013. In fact, there was a trend of growth for the exchange rate with certain fluctuations.

\noindent \textbf{3.~Exchange rate in the monetary model of money circulation}

The Ukrainian history of the exchange rate is very short, for it has always been tied to the US dollar. Recently, it has indeed become the exchange rate, that should counterbalance the trade balance. It is important to understand the theoretical and practical basis for its formation. Economists [8, 9, 10, 11] made an important contribution to understanding the formation of exchange rates, but unsolved problems remain. In [12, 13, 14], a method to manage the hryvnia / dollar exchange rate was proposed.

  The monetary approach is based on the postulate of purchasing power parity. This model can not predict the exchange rate more accurately than the random walk model. In view of the importance of the approach, that until now remains a paradigm for the formation of the exchange rate, we will check it using the example of the hryvnia / dollar exchange rate formation. The monetary approach proceeds from the definition of the exchange rate as a relative price of money and tries to model this relationship through supply and demand for money.

  Consider discrete evolution of the hryvnia / dollar exchange rate. If${}_{} $\textbf{$p_{t} $ }and $p_{t}^{*} $ are logarithms of price levels in the time moment $t$, then up to the constant depending only on the choice of the base year prices, $p_{t} -p_{t}^{*} $ is logarithm of the hryvnia / dollar exchange rate. To obtain the equation managing the exchange rate, take the money circulation equations for two countries
\begin{equation} \label{GrindEQ__8_} 
m_{t} =p_{t} +ky_{t} -\lambda i_{t} ,                                                                        
\end{equation} 
\begin{equation} \label{GrindEQ__9_} 
m_{t}^{*} =p_{t}^{*} +k^{*} y_{t}^{*} -\lambda ^{*} i_{t}^{*} ,                                                                    
\end{equation} 
where $m_{t} ,{\kern 1pt} \, y_{t} ,\; i_{t} \, $are logarithms of money demand, gross income and interest rate at the moment $t,$ and $k,\; \lambda $ are positive constants. Eq. \eqref{GrindEQ__9_} refers to a foreign economy. Given that the exchange rate                              
\begin{equation} \label{GrindEQ__10_} 
s_{t} =p_{t} -p_{t}^{*} +a ,                                                                        
\end{equation} 
and the formulae \eqref{GrindEQ__8_}, \eqref{GrindEQ__9_}, we have
\begin{equation} \label{GrindEQ__11_} 
s_{t} =a+m_{t} -m_{t}^{*} -(ky-k^{*} y^{*} )+\lambda i_{t} -\lambda ^{*} i_{t}^{*}  .                                      
\end{equation} 

Equation \eqref{GrindEQ__11_} requires many assumptions. Among them, the flexibility of prices in both economies. Argumenting neglect of the market for goods, labor, foreign economic relations, and bonds is difficult. One simplifies the model assuming elastic income, and the same rate for domestic and foreign economies, implying that
\begin{equation} \label{GrindEQ__12_} 
s_{t} =a+m_{t} -m_{t}^{*} -k(y-y^{*} )+\lambda (i_{t} -i_{t}^{*} ) .                                       
\end{equation} 

 The coefficients entering into Eq.  \eqref{GrindEQ__12_} must be determined.

\noindent \textbf{4.~Exchange rate and inflation targeting model }

Simple relations between the exchange rate, the price level, the real product, the demand for money, and the price of them from the equation of monetary circulation can not hold deterministically in view of the number of restrictions for the monetary model to be valid. Just the monetary circulation equation itself is not known.

~~ Further, we assume that the deviation of the left side of \eqref{GrindEQ__12_} from the right one is a random process.

  Therefore, let discrete random process $\varsigma _{k+1,i} ,\; i=\overline{1,n,}$ exists and $k$ such factors  $X_{i} =\{ x_{ji} \} _{j=0}^{k} ,\; i=\overline{0,n,}$ $X_{0} =\{ x_{j0} \} _{j=1}^{k} ,\; x_{j0} =1,\; j=\overline{1,k},$ that
\begin{equation} \label{GrindEQ__13_} 
\sum _{s=0}^{p}a_{s} \varsigma _{k+1,i-s}  -f_{i} (X_{0} ,X_{1} ,...,X_{k} )=\varepsilon _{i} ,\quad i=\overline{1,n,} 
\end{equation} 
where $a_{0} =1,$ are random values, $\varsigma _{k+1,-p} ,${\dots},$\varsigma _{k+1,0} $ are known, and random values $\varepsilon _{i} $, $i=\overline{1,n,}$ are independent random values with zero mean and dispersion $\sigma ^{2} $, also, $\varsigma _{k+1,-p} ,${\dots},$\varsigma _{k+1,0} $ do not depend on $\varepsilon _{i} $, $i=\overline{1,n.}$ Let functions $f_{i} (X_{0} ,X_{1} ,...,X_{k} )$be nonlinear functions of the factors. The most important case is when 
\[f_{i} (X_{0} ,X_{1} ,...,X_{k} )=\sum _{j=0}^{k}b_{j} x_{ji}  ,\; \quad i=\overline{1,n,}\] 
and equalities  \eqref{GrindEQ__13_} become
\begin{equation} \label{GrindEQ__14_} 
\varsigma _{k+1,i} -\sum _{j=0}^{k}b_{j} x_{ji}  =\varepsilon _{i} ,\quad i=\overline{1,n.} 
\end{equation} 

Therefore, we further assume that the set of random variables $\varsigma _{k+1,i} ,\; i=\overline{1,n,}$ is independent and has joint normal distribution with mean $E\varsigma _{k+1,i} =\sum _{j=0}^{k}b_{j} x_{ji}  $and dispersion $\sigma ^{2} $. It follows from  \eqref{GrindEQ__14_} that the best forecast of a random process $\varsigma _{k+1,i} ,\; i=\overline{1,n,}$ is its mean. To estimate the maximum likelihood and the determination coefficient, which is the correlation of the process and its prediction, we introduce the matrix $X,$ whose columns are the vectors $X_{i} ,\; i=\overline{0,k},$ where
\begin{equation} \label{GrindEQ__15_} 
X_{i} =\{ x_{ij} \} _{j=1}^{n} ,\; \; \; i=\overline{1,k,} X_{0} =\{ e_{j} \} _{j=1}^{n} ,\; \; e_{j} =1,\; \; j=\overline{1,n.} 
\end{equation} 

Then this matrix can be represented as
\[X=\left(\begin{array}{l} {1,x_{11} ,...,x_{k1} } \\ {.................} \\ {1,x_{1n} ,...,x_{kn} } \end{array}\right),\] 
and the set of eqs  \eqref{GrindEQ__14_} as
\begin{equation} \label{GrindEQ__16_} 
\varsigma _{k+1} =Xb+\varepsilon , 
\end{equation} 
where $\varsigma _{k+1} =\{ \varsigma _{k+1,i} \} _{i=1}^{n} $, $\varepsilon =\{ \varepsilon _{i} \} _{i=1}^{n} ,$are column vectors containing corresponding elements $\varsigma _{k+1,i} $ and $\varepsilon _{i} $. Let matrix $X$ rank is $k+1.$ Let $A=X^{T} X,$ it is symmetrical matrix having inverse one and the maximum likelihood estimate (MLE) for the coefficients of the column vector $b=\{ b_{0} ,b_{1} ,...,b_{k} \} $ is unbiased and can be given in the form 
\begin{equation} \label{GrindEQ__17_} 
\overline{b}=A^{-1} X^{T} \varsigma _{k+1} .                                                                                        
\end{equation} 

The quality of the regression model is determined by the selective multiple correlation coefficient between the sample and the forecast
\begin{equation} \label{GrindEQ__18_} 
R_{n-1} =\frac{\sum _{i=1}^{n}(x_{k+1,i} -\overline{x_{k+1} })(x_{k+1,i}^{1} -\overline{x_{k+1}^{1} }) }{\sqrt{\sum _{i=1}^{n}(x_{k+1,i} -\overline{x_{k+1} })^{2} \sum _{i=1}^{n}(x_{k+1,i}^{1} -\overline{x_{k+1}^{1} }  } )^{2} } ,                                                    
\end{equation} 

where $x_{k+1}^{1} =X\overline{b}=\{ x_{k+1,i}^{1} \} _{i=1}^{n} $, 
and $\overline{x_{k+1} }=\frac{\sum _{i=1}^{n}x_{k+1,i}  }{n} $, $\overline{x_{k+1}^{1} }=\frac{\sum _{i=1}^{n}x_{k+1,i}^{1}  }{n} $.

It is easy to show that for $R_{n-1}^{2} $the representation
\begin{equation} \label{GrindEQ__19_} 
R_{n-1}^{2} =\frac{\sum _{i=1}^{n}(x_{k+1,i}^{1} -\overline{x_{k+1}^{1} } )^{2} }{\sum _{i=1}^{n}(x_{k+1,i} -\overline{x_{r+1} })^{2}  }  
\end{equation} 
is valid. $R_{n-1}^{2} $ is called determination coefficient. The closer it is to unity, the better the quality of regression. For calculations, it is convenient to represent the determination coefficient in the form 
\begin{equation} \label{GrindEQ__20_} 
R_{n-1}^{2} =\frac{<x_{k+1} -\overline{x_{k+1} },X_{1} A_{1}^{-1} X_{1}^{T} (x_{k+1} -\overline{x_{k+1} })>}{<x_{k+1} -\overline{x_{k+1} },x_{k+1} -\overline{x_{k+1} }>}  ,                                                   
\end{equation} 
where
\begin{equation} \label{GrindEQ__21_} 
X_{1} =\left(\begin{array}{l} {x_{11}^{1} ,...,x_{k1}^{1} } \\ {.................} \\ {x_{1n}^{1} ,...,x_{kn}^{1} } \end{array}\right),                                                                                   
\end{equation} 
$x_{ji}^{1} =x_{ji} -\frac{\sum _{i=1}^{n}x_{ji}  }{n} $, $A_{1} =X_{1}^{T} X_{1} $, $x_{k+1} -\overline{x_{k+1} }=\{ x_{k+1,i} -$$\frac{\sum _{i=1}^{n}x_{k+1,i}  }{n} $$\mathrm{\}}$${}_{i=1}^{n} $, and $<a,b>$is scalar product of the vectors $a$and $b.$ Estimation of maximum likelihood for dispersion $\sigma ^{2} $ is
\begin{equation} \label{GrindEQ__22_} 
\sigma ^{2} =\frac{<x_{k+1} -\overline{b},x_{k+1} -\overline{b}>}{n} . 
\end{equation} 

One can judge the quality of the factors on the basis of the Snedekor-Fisher statistics 
\begin{equation} \label{GrindEQ__23_} 
\eta _{n-k-1,k+1} =\frac{<\varepsilon ,(I-XA^{-1} X^{T} )\varepsilon >}{<\varepsilon ,XA^{-1} X^{T} \varepsilon >} \frac{(k+1)}{(n-k-1)} , 
\end{equation} 
where$\varepsilon =\{ \varepsilon _{i} \} _{i=1}^{n} $, and random values $\varepsilon _{i} =\frac{\varsigma _{k+1,i} -E\varsigma _{k+1,i} }{\sigma } ,\; \; i=\overline{1,n},$ have normal distribution $N(0,1)$ with zero mean and unit dispersion and are reciprocally independent. $E\varsigma _{k+1,i} $ is mean value of the random value $\varsigma _{k+1,i} .$

\noindent \textbf{5.~Money circulation equation for the Ukrainian economy in 2012-2014}

  In this section, using the statistics of 2012-2014, we establish a stochastic difference equation that hryvnia / dollar exchange rate satisfies. Since the quarterly statistical data 2012-2014 are known, in our case we use the following notation. Dependent variable $Y_{i} ,\; i=\overline{1,n,}$ describes hryvnia / dollar exchange rate evolution. Introduce the vector $Y=\{ Y_{i} \} _{i=1}^{n} .$ Let $X_{0} =\{ e_{i} \} _{i=1}^{n} ,\; \; e_{i} =1,\; \; i=\overline{1,n,}$ is unit vector-column. Let $G_{1} =\{ G_{i}^{1} \} _{i=1}^{n} $and$G_{2} =\{ G_{i}^{2} \} _{i=1}^{n} $ are vectors-columns of gross domestic products for Ukraine and US, resp., and$ $$M_{1} =\{ M_{i}^{1} \} _{i=1}^{n} ,$ $M_{2} =\{ M_{i}^{2} \} _{i=1}^{n} $ are money supply vectors for Ujraine and US. Let $R_{1} =\{ R_{i}^{1} \} _{i=1}^{n} ,$ $R_{2} =\{ R_{i}^{2} \} _{i=1}^{n} $ are refinancing rates of the National Bank and the US Federal Reserve System. Introduce the vectors 
\[m_{1} =\log (M_{1} )=\{ \log (M_{i}^{1} )\} _{i=1}^{n} =\{ m_{i}^{1} )\} _{i=1}^{n} , m_{2} =\log (Y*M_{2} )=\{ \log (Y_{i} M_{i}^{2} )\} _{i=1}^{n} =\{ m_{i}^{2} )\} _{i=1}^{n} ,\] 
\[r_{1} =\log (R_{1} )=\{ \log (R_{i}^{1} )\} _{i=1}^{n} =\{ r_{i}^{1} \} _{i=1}^{n} , r_{2} =\log (R_{2} )=\{ \log (R_{i}^{2} )\} _{i=1}^{n} =\{ r_{i}^{2} \} _{i=1}^{n} ,\] 
\[g_{1} =\log (G_{1} )=\{ \log (G_{i}^{1} )\} _{i=1}^{n} ==\{ g_{i}^{1} \} _{i=1}^{n} , g_{2} =\log (Y*G_{2} )=\{ \log (Y_{i} G_{i}^{2} )\} _{i=1}^{n} =\{ g_{i}^{2} \} _{i=1}^{n} ,\] 
\[y=\log (Y)=\{ \log (Y_{i} )\} _{i=1}^{n} =\{ y_{i} \} _{i=1}^{n} .\] 

Let vector $y_{i} $be random process satisfying stochastic equation
\begin{equation} \label{GrindEQ__24_} 
y_{i} =b_{0} +b_{1} (m_{i}^{1} -m_{i}^{2} )+b_{2} (g_{i}^{1} -g_{i}^{2} )+b_{3} (r_{i}^{1} -r_{i}^{2} )+\varepsilon _{i} ,  i=\overline{1,n}, 
\end{equation} 
where $\varepsilon _{i} $ are independent random values distributed by the law $N(0,\sigma ).$Introduce vector-columns$X_{1} =$$m_{1} -m_{2} $, $X_{2} =$$g_{1} -g_{2} ,$$X_{3} =$$r_{1} -r_{2} $, where $X_{i} =\{ x_{ij} \} _{j=1}^{n} ,\; i=\overline{0,3.}$ We have the problem of parameter estimation as in the problem \eqref{GrindEQ__14_} with $b_{1} =1,$ and $b_{2} <0,\; b_{3} >0.$ The maximum likelihood estimate gives $b_{0} =0.67707,\; $$b_{2} =-1.15037,$$b_{3} =0.39375.$ Determination coefficient$R_{11}^{2} =0.76750,$ and dispersion estimation $\sigma =0.031653.$ Assuming that the exchange rate is the ratio of price levels in Ukraine and the United States, one can, to within a constant multiplier, write the equation of money circulation
\begin{equation} \label{GrindEQ__25_} 
M_{1} =P_{1} G_{1}^{-b_{2} } R_{1}^{-b_{3} } ,                                                                
\end{equation} 
where $M_{1} $ is money demand, $G_{1} $ is nominal GDP, $R_{1} $ is refinancing rate of the National Bank, $P_{1} $ is price level. If the rate of money circulation $V_{1} $, then
\begin{equation} \label{GrindEQ__26_} 
M_{1} =\frac{G_{1} }{V_{1} } .                                                                                   
\end{equation} 

Equating \eqref{GrindEQ__25_} and \eqref{GrindEQ__26_}, we have the rate of money circulation in the form
\begin{equation} \label{GrindEQ__27_} 
V_{1} =\frac{R_{1}^{b_{3} } }{P_{1} G_{1}^{-b_{2} -1} } .                                                                             
\end{equation} 

The calculation of the money circulation rate shows the growing dynamics, i.e., the unwinding of inflation and the fall in the standard of living.

\noindent \textbf{6.~Regression equation for the exchange rate and its forecast}

\noindent 

The obtained estimates in Eq. \eqref{GrindEQ__24_} allow us to predict the exchange rate. But the low coefficient of determination stimulates the delay. Let the random process $y$ satisfy the stochastic equation 
\begin{equation} \label{GrindEQ__28_} 
y_{i} =b_{0} +b_{1} (m_{i}^{1} -m_{i}^{2} )+b_{2} (g_{i}^{1} -g_{i}^{2} )+b_{3} (r_{i}^{1} -r_{i}^{2} )+b_{4} y_{i-1} +\varepsilon _{i} ,  i=\overline{1,n}, 
\end{equation} 
where $\varepsilon _{i} $ are independent random values distributed by the law $N(0,\sigma )$, and $y_{0} $is known number. As earlier, introduce vectors-columns $X_{1} =$$m_{1} -m_{2} $, $X_{2} =$$g_{1} -g_{2} ,$$X_{3} =$$r_{1} -r_{2} $, $X_{4} =\{ y_{i-1} \} _{i=1}^{n-1} $, where $X_{i} =\{ x_{ij} \} _{j=1}^{n} ,\; i=\overline{0,4.}$ We have parameter estimate problem as in \eqref{GrindEQ__14_} with $b_{1} =1,$ and $b_{2} <0,\; b_{3} >0.$ Least squares estimation gives $b_{0} =-1.61447,$ $b_{2} =-0.46690,$ $b_{3} =0.21907,$ $b_{4} =1.53031$ for $y_{0} =2.068.$ Determination coefficient $R_{11}^{2} =0.95280,$ and dispersion estimate $\sigma =0.0070095.$ The assumption that the exchange rate is a purchasing power parity, up to a constant factor, gives the following recurrence relation
\begin{equation} \label{GrindEQ__29_} 
P_{i} =P_{i-1}^{b_{4} } M_{i} G_{i}^{b_{2} } R_{i}^{b_{3} } ,  i=\overline{2,n}. 
\end{equation} 

Having $M_{i} =\frac{G_{i} }{V_{i} } $, we obtain
\begin{equation} \label{GrindEQ__30_} 
V_{i} =P_{i-1}^{b_{4} } P_{i}^{-1} G_{i}^{b_{2} +1} R_{i}^{b_{3} } .                                                                         
\end{equation} 

And in this case, the pronounced growing dynamics of the money circulation rate is confirmed, showing the untwisting of inflation. The best forecast for the exchange rate is the expression
\begin{equation} \label{GrindEQ__31_} 
b_{0} +b_{1} (m_{i}^{1} -m_{i}^{2} )+b_{2} (g_{i}^{1} -g_{i}^{2} )+b_{3} (r_{i}^{1} -r_{i}^{2} )+b_{4} y_{n-1} . 
\end{equation}

\noindent \textbf{7.~Dependence of the exchange rate on internal factors}

\noindent \textbf{}

We'll find out the effect of inflation, the budget deficit, the level of energy prices, and money supply on the exchange rate. As before, we assume the exchange rate to be a random process that satisfies the system of equations \eqref{GrindEQ__13_}. More detailed, the exchange rate is influenced by factors $X_{i} =\{ x_{ij} \} _{j=1}^{n} ,\; i=\overline{0,4,}$ where $X_{0} =\{ e_{i} \} _{i=1}^{n} ,\; \; e_{i} =1,\; \; i=\overline{1,n,}$ is unit vector-column, $X_{1} =\{ x_{1j} \} _{j=1}^{n} $is inflation level, $X_{2} =\{ x_{2j} \} _{j=1}^{n} $is state deficit, $X_{3} =\{ x_{3j} \} _{j=1}^{n} $energy carriers' price, $X_{4} =\{ x_{4j} \} _{j=1}^{n} $is money supply. With $y=\log (Y)=\{ \log (Y_{i} )\} _{i=1}^{n} =\{ y_{i} \} _{i=1}^{n} $, we'll assume random process $y$ satisfy the set of equations
\begin{equation} \label{GrindEQ__32_} 
y_{i} =b_{0} +b_{1} x_{1i} +b_{2} x_{2i} +b_{3} x_{3i} +b_{4} x_{4i} +b_{5} y_{i-1} +\varepsilon _{i} ,  i=\overline{1,n}, 
\end{equation} 
where $\varepsilon _{i} ,\; \; i=\overline{1,n},$ is the set of independent distributed by normal law $N(0,\sigma )$ random values, and $y_{0} $is fixed number. Maximum likelihood estimate for the vector $B=\{ b_{i} \} _{i=0}^{5} $ gave the result $b_{0} =3.8,$ $b_{1} =3.3,$ $b_{2} =-1.4{\kern 1pt} \cdot \; 10^{-7} ,$ $b_{3} =5.1\cdot 10^{-3} $, $b_{4} =4.7\cdot 10^{-8} $, $b_{5} =-1.6\cdot 10^{-1} $ for $y_{0} =6.68.$ Determination coefficient $R_{11}^{2} =0.99,$ and dispersion estimate $\sigma =4.1\cdot 10^{-4} $. The latter indicates a very close relationship of the exchange rate with the level of inflation, the deficit of the state budget, the level of energy prices, and the supply of money. Therefore, to stabilize the exchange rate, a balanced policy is needed to reduce the deficit of the state budget, energy intensity of production and household consumption of energy resources. The National Bank should keep inflation balanced by monetary policy. This is a policy of reducing the refinancing rate, maintaining the liquidity of the banking system with a moderate emission of the national currency.

\noindent 

\noindent \textbf{8.~Inflation Targeting}

\noindent \textbf{}

The fact the National Bank should keep inflation is natural. This is in keeping with the spirit of the federal reserve system. We will find out what factors influence inflation in Ukraine.

~ Let the level of inflation be a random process satisfying the set of equations \eqref{GrindEQ__13_}. The level of inflation is influenced by factors $X_{i} =\{ x_{ij} \} _{j=1}^{n} ,\; i=\overline{0,3,}$ where $X_{0} =\{ e_{i} \} _{i=1}^{n} ,\; \; e_{i} =1,\; \; i=\overline{1,n,}$ is unit vector-column, $X_{1} =\{ x_{1j} \} _{j=1}^{n} $is exchange rate, $X_{2} =\{ x_{2j} \} _{j=1}^{n} $is state deficit, $X_{3} =\{ x_{3j} \} _{j=1}^{n} $is trade balance deficit. Let $Y=\{ Y_{i} \} _{i=1}^{n} $ be inflation level, and a random process $y=\log (Y)=\{ \log (Y_{i} )\} _{i=1}^{n} =\{ y_{i} \} _{i=1}^{n} $satisfy the set of equations
\begin{equation} \label{GrindEQ__33_} 
y_{i} =b_{0} +b_{1} x_{1i} +b_{2} x_{2i} +b_{3} x_{3i} +\varepsilon _{i} ,  i=\overline{1,n}, 
\end{equation} 
where $\varepsilon _{i} ,\; \; i=\overline{1,n},$ is the set of independent distributed by the normal law $N(0,1)$ random values. Maximum likelihood estimate for the vector $B=\{ b_{i} \} _{i=0}^{4} $ gave the result $b_{0} =-2.6\cdot 10^{-1} ,$ $b_{1} =3.3\cdot 10^{-4} ,$ $b_{2} =-4.7{\kern 1pt} \cdot \; 10^{-8} ,$ $b_{3} =-1.8\cdot 10^{-7} $. Determination coefficient $R_{11}^{2} =0.99,$ and dispersion etimate $\sigma =3.5\cdot 10^{-5} .$ Therefore, inflation level depends on state deficit and trade balance deficit.

\noindent 

\noindent \textbf{9.~Conclusion}

The stochastic equation for an adequate hryvnia / dollar and inflation rates is obtained. It has been established that the hryvnia / dollar course was artificially maintained during 2012--2013. We describe the hryvnia / dollar exchange rate evolution by a random process satisfying the difference stochastic equation. We confirm the observation that the logarithm of the exchange rate is a random walk without drift. The factors controlling the process are established. On this basis, the money circulation equation is found for 2012--2014. An increase in the rate of money circulation is obtained for that time indicating a fall in the standard of living. A close positive correlation of the hryvnia / dollar exchange rate with that of inflation, a deficit in the state budget, energy prices, and the supply of money has been found. Recommendations are made for macroeconomic policies to stabilize the Ukrainian economy. In the process of ascertaining what a macroeconomic policy should be to reduce inflation, a close positive correlation of the inflation rate with the hryvnia / dollar exchange rate, state deficit, and a trade balance deficit was established. That is, it is impossible to stabilize inflation without stabilizing the exchange rate.

\newpage

\noindent \textbf{References}

\noindent 1. R.E. Kalman, A new approach to linear filtering and prediction problems, Transactions of the ASME, Ser.D, Journal of Basic Engineering, 82, 34 -- 45, (1960).

\noindent 2. Kalman Filtering and Neural Networks, edited by Simon Haykin, New York, John Willey \& Sons, Inc., 2001, 410p.

\noindent 3.~M.L. Mussa, The exchange rate determination, Volume URL: http:/www.nbr:org/books/bils84-1, 13-78, (1984).

\noindent 4.~N.S. Gonchar, A.S. Zhokhin. Critical States in Dynamical Exchange Model and Recession Phenomenon, 

\noindent Journal of Automation and Information Scince. \textbf{45,} 50-58 (2013).

\noindent 5.~Gonchar N.S. Mathematical foundations of information economics. - Kiev: Inst. for Theoretical Physics, 

\noindent     2008. -- 468p.

\noindent 6.~N.S.Gonchar, A.S.Zhokhin, W.H.Kozyrski. On Mechanism of Recession Phenomenon. Journal of Auto-

\noindent mation and Information Sciences, 47, 4, p. 1-17 (2015)

\noindent 7.~N.S. Gonchar, A.S. Zhokhin, W.H. Kozyrski, General Equilibrium and Recession Phenomenon, American

\noindent     Journal of Economics, Finance and Management. \textbf{1,} 559-573 (2015).

\noindent 8.~B. Balassa, The Purchasing Power Parity: A Reappraisal. Journal of Political Economy,72, 584-596 (1964)

\noindent 9.~R. Dornbusch, Exchange rate dynamics: where do we stand? Brookings Papers on Economic Activity, 143-

\noindent     155, (1980).

\noindent 10.~R.A. Meese, K. Rogoff. Empirical Exchange rate models of the seventies. Journal of International 

\noindent     Economics. Do they fit out of sample? \textbf{14,} 3-24, (1983). 

\noindent 11.~C.J. Neely, L. Sarno, How well do Monetary fundamentals forecast exchange rates? The Federal Reserve 

\noindent     Bank of St. Louis, (2002)

\noindent 12. V.D. Romanenko, O.A. Reutov, Optimal decisions adoption to stabilize hryvnia / dollar cource based on mathematical models with varying discretization degrees. Science news of NTUU KPI, 6, p. 67-73, (2011).

\noindent 13. V.D. Romanenko, O.A. Reutov, Modeling and optimal decision-making for maintenance the stability of the consumer price index. System research and Information Technologies, 4, p. 23-34 (2012).

\noindent 14. V.D. Romanenko, O.A. Reutov, Systematization of mathematical models with varying discretization degrees~for the dynamic processes of financial infrastructure. System research and information technologies, 2, p. 54-66, (2013).

\textbf{}

\end{document}